\documentstyle[11pt,aaspp4]{article}

\tighten

\input psfig

\def\kmsm{\,{\rm km\, s^{-1}\, Mpc^{-1}} }

\begin{document}

\title{Constraining Galaxy Mass Profiles with Strong Gravitational Lensing}
\author{David Rusin} 
\affil{Department of Physics and Astronomy, University of Pennsylvania, 
209 S. 33rd St., Philadelphia, PA, 19104-6396}

\begin{abstract}

Measured time delays between the images of a gravitationally lensed source can
lead to a determination of the Hubble constant ($H_o$), but only if the
lensing mass distribution is well understood. The inability to sufficiently
constrain galaxy mass models results in large uncertainties on the derived
$H_o$, and severely hampers the cosmological application of this otherwise
elegant method. At the same time, lensing must compete with new techniques
that have the potential to measure the Hubble constant to within a few percent
by the middle of the decade. In this letter we re-evaluate the role of strong
gravitational lensing in the age of precision cosmology, and present a series
of Monte Carlo simulations that demonstrate the effect of the galaxy mass
distribution on Hubble constant determination.  Though most gravitational lens
systems are unlikely to contribute significantly to precision studies of the
cosmological distance scale, the strong dependence of predicted time delays on
the galaxy mass profile suggests that the most useful and interesting results
may be obtained by running the traditional lensing problem in reverse --
namely, combining measured time delays with a well-constrained value of $H_o$
to study galaxy mass distributions.

\end{abstract}


\section{Introduction} 

Gravitational lensing offers a potent tool for determining the cosmological
distance scale (Refsdal 1964), but only if the mass distribution of the
deflector is well constrained. Since the mass model predicts time delays
between the lensed images up to a factor of $h^{-1}$, where $H_o = 100h\kmsm$,
a simple comparison with measured time delays can yield the Hubble
constant. This method is particularly tempting as it would determine $H_o$
directly out to cosmological distances, forgoing the calibration errors
associated with traditional distance-ladder techniques (e.\ g.\ Mould et al.\
2000).  Much of the recent interest in finding arcsecond-scale gravitational
lens systems through large, dedicated programs such as the Cosmic Lens All-Sky
Survey (CLASS; Myers et al.\ 2000) has been motivated by this goal.

Measurements of $H_o$ from gravitational lenses are only as good as our
understanding of the galaxies doing the lensing, as the model-predicted time
delays contribute an equal fractional uncertainty to the derived Hubble
parameter. Unfortunately, experience has shown that our knowledge of lensing
mass distributions is not very good at all. This problem has been greatly
exacerbated by the observation that a significant fraction of lenses,
including several of those with measured time delays, are quite complex in
structure (e.\ g.\ CLASS B1359+154, Rusin et al.\ 2000; QSO 0957+561, Barkana
et al.\ 1999; PG 1115+080, Schechter et al.\ 1997; CLASS B1608+656, Koopmans
and Fassnacht 1999). Mass models are difficult to construct for such systems
due to limited constraints and parameter degeneracies (e.\ g.\ the mass-sheet
degeneracy, Gorenstein et al.\ 1988).

Significant uncertainties in the mass model can remain even in systems lensed
by a single isolated galaxy. This is because the positions and flux density
ratios of the lensed images constrain derivatives of the potential, rather
than the potential itself. Often a range of models can provide sufficient fits
to the lens system while placing different potentials at the image
positions. The predicted time delays will be greatly affected by model
degeneracies, as they depend on the potential directly (Schneider, Ehlers and
Falco 1992). This is particularly discouraging for many two-image lens
systems, which do not offer a sufficient number of constraints to even attempt
fits to the profile shape. For such systems the profile must be assumed, and
is commonly taken to be isothermal. The uncertainty in the slope of the mass
profile and its effect on Hubble constant determination are rarely included in
most preliminary modeling analyses.

Attempts to constrain galaxy mass profiles through strong lensing generally
focus on image positions, flux density ratios and, if available, the ratios of
measured time delays. Studies of this type have thus far been performed in
only a small handful of lens systems, and have had limited success in pinning
down the mass profile. Assuming a power-law surface mass density ($\Sigma \sim
r^{-\gamma}$), it has been found that the merging lens galaxies of CLASS
B1608+656 are best fit by $0.8 \leq \gamma \leq 1.2$ (Koopmans and Fassnacht
1999), while the large elliptical galaxy in QSO 0957+561 is best fit by $0.82
\leq \gamma \leq 0.93$ (Grogin and Narayan 1996). However, each of these
systems is affected by complicated deflectors. A profile analysis has also
been done for MG 1654+134 (Kochanek 1995), which is a radio ring system lensed
by an isolated galaxy. Because the lensed extended emission probes a wide
range of radii (as opposed to the images of a compact source), ring systems
can produce tighter constraints on the mass profile ($0.9 \leq \gamma \leq
1.1$ for MG 1654+134).

While isothermal ($\gamma=1$) mass profiles are largely compatible with
lensing, a range of near-isothermal profiles ($0.8 \leq \gamma \leq 1.2$) are
equally feasible. In this Letter I outline the systematic effects of this
uncertainty on the predicted time delays, and evaluate the possibility of
measuring $H_o$ with the set of lens systems currently known. I then consider
the largely-ignored reverse problem in lensing -- namely, using measured time
delays along with a well-determined value of $H_o$ to study galaxy mass
profiles. The results offered by this technique have the potential to provide
greatly improved constraints on the inner mass distribution for galaxies at
intermediate redshift.

\section{Power-law mass distributions}

In this Letter I approximate the lensing mass of galaxies by a singular
power-law ellipsoid (SPLE), with surface mass density
$$\Sigma(x_1, x_2) = q [x_1^2 + f^2 x_2^2]^{-\gamma/2}$$ where $\gamma$ is
radial profile, $f$ is the axial ratio and $q$ is the normalization. Only
recently have power-law mass profiles for arbitrary $\gamma$ become a feasible
alternative for studying and modeling gravitational lens systems. All
calculations presented in this paper make use of the rapidly converging series
solutions for the deflection angles and magnification matrices derived by
Barkana (1998) and implemented in the ``FASTELL'' software package.

While even small core radii greatly affect lensing cross-sections, they do not
significantly alter the properties of the images in a given system and can
therefore be ignored in this analysis. The SPLE probably oversimplifies the
overall mass distribution in galaxies, but may be a feasible approximation
when limited to describing the inner few $h^{-1}$ kpc that is probed by strong
lensing. Moreover, it has been shown that a large number of gravitational lens
systems can be successfully modeled using power-law mass profiles, isothermal
or otherwise.

The lensing properties of the singular power-law ellipsoid depend strongly on
the value of the profile slope $\gamma$.  Here we summarize the possible image
configurations produced by the SPLE for the range of profiles ($0.7 < \gamma <
1.3$) and axial ratios ($f > 0.5$) that seem to be favored by most
gravitational lens systems:

\begin{itemize}

\item For the familiar isothermal case ($\gamma=1$; Kormann, Schneider and
Bartelmann 1994a), the SPLE will produce one, two or four images, as is
typically observed in gravitational lens systems. The third image (for
doubles) and fifth image (for quads) does not exist, as the radial critical
curve that would enclose this image degenerates to a point at the lens center. 

\item For profiles steeper than isothermal ($\gamma>1$), the deflection angle
diverges as $r \rightarrow 0$, so the SPLE will multiply-image all source
positions, producing either two or four images. The lensing cross-section for
$\gamma>1$ is thus nominally infinite. However, the second image is highly
demagnified if the source is sufficiently misaligned with the lens, which
leads to a large flux ratio between the images. This results in a finite
``observational cross-section'' that depends on the dynamic range limits of a
given survey. The CLASS survey, for example, can only identify lens systems in
which the flux density ratio between the brightest images is $\leq$ 10:1.

\item For profiles shallower than isothermal ($\gamma<1$), the radial critical
curve exists, allowing for the creation of one, three or five images.  The
third image and fifth image will be sufficiently demagnified to escape
detection if $\gamma$ is close to 1. For shallower profiles, however, 
the SPLE commonly produces two distinct classes
of three-image systems that would be readily observable -- (1) fold-imaged
configurations in which the positive parity image enclosed by the radial
critical curve is sufficiently bright (3F-type; Fig.\ 1a), and (2) naked-cusp
configurations in which a source enclosed by the diamond caustic only is
lensed into three bright images on the same side of the lens, with no counter
images (3C-type; Fig.\ 1b). The paucity of observational evidence for
three-image systems suggests that the typical mass profile of lens
galaxies is unlikely to be much shallower than $\gamma = 0.75$ (Rusin and Ma
2000).

\end{itemize}

\section{Mass profiles and time delays}

How well can we hope to constrain galaxy mass profiles using gravitational
lensing? How does the remaining uncertainty affect the predicted time delays,
and therefore the extracted Hubble constant? Analyses of this type have thus
far been been performed for only a few specific lenses, and in many instances
have been significantly complicated by the presence a complex deflector
system. Here I consider the case of lens systems produced by a single isolated
galaxy, and address the above questions for both two and four-image
morphologies. 

\subsection{Two-image systems}

Gravitational lenses consisting of two unresolved images comprise the majority
of known lens systems. Such systems provide only five constraints to the lens
model (four image coordinates and the flux density ratio). Even if the mass
distribution can be fixed to a well known galaxy position, one must still
solve for the axial ratio, position angle and normalization of the model, as
well as the two coordinates of the unlensed source. Because there are no
degrees of freedom, any mass profile can fit the data perfectly ($\chi^2 =
0$). Such lens systems are thus not capable of constraining the galaxy mass
profile, if only image positions and flux density ratios are used.

To study the effect of the mass profile on the predicted time delays, a series
of fake two-image lens systems was generated using a range of SPLE mass models
with $0.7 \leq \gamma \leq 1.3$ and $ 0.5 \leq f \leq 0.9$.  To limit the
analysis to the types of lens systems that are actually observed,
configurations with a flux density ratio of $\geq$ 10:1 between the two
brightest images or with an observable third image (flux density ratio between
the brightest and third brightest image of $\leq$ 50:1) were discarded.  Next,
the fake lens systems were remodeled using an isothermal ($\gamma=1$) mass
profile. The mass distribution was fixed at the same position as the SPLE to
constrain the model. Finally, the time delays of the isothermal and power-law
distributions were compared.

The predicted time delay relative to the isothermal case ($\Delta
t_{pred}(\gamma) / \Delta t_{pred}(\gamma=1)$) is displayed in Fig.\
2.\footnote{The results described here were originally presented at
the 195th AAS Meeting in January. Since then, an interesting paper has been
submitted by Witt, Mao and Keeton (2000), which references a similar set of
simulations describing the effect of the galaxy mass profile on time
delays. These results agree well with those shown in Fig.\ 2.}. The error bars
indicate the range containing 95\% of the Monte Carlo trials. The scatter in
the results is remarkably small, given that the lens systems were created by
placing sources at random locations behind the galaxy. The plot is shown for
SPLEs with $f=0.7$, but the results are only dependent on axial ratio at the
few percent level.  The results are also independent of the angular separation
of the gravitational lens system. (A simple calculation shows that the size of
the caustics and critical curves scale as $q^{1/\gamma}$, and the absolute
time delays scale as $q^{2/\gamma}$.)

For a two-image lens system, the predicted time delays vary almost linearly
with $\gamma$, as found independently by Witt, Mao and Keeton (2000). Steeper
mass profiles lead to longer predicted delays, since the ray paths traverse 
deeper potentials. Similar trends have been noted in several previous
investigations (Barkana et al.\ 1999, Koopmans and Fassnacht 1999, Williams
and Saha 2000). A longer predicted delay translates into a larger value of
the Hubble constant.  In almost all trials, the position angle of the SIE was
recovered to within a few degrees of that of the original SPLE. The axial
ratio required to fit the lens system decreases (becomes flatter) as the
profile is made steeper. However, since there is no reason to assume that the
axial ratio of the mass distribution must match that of the observed surface
brightness distribution of the lensing galaxy, this does not provide a useful
constraint on the model.

The range of profiles consistent with lensing ($0.8 \leq \gamma \leq 1.2$)
translates into a spread of $\sim 20\%$ in the predicted time delays. This
uncertainty should be folded into preliminary modeling analyses of all
two-image lenses. It is therefore doubtful whether any basic two-image
gravitational lens system will produce a precision measurement of the Hubble
constant.

In a fraction of radio-loud gravitational lens systems, VLBI imaging is able
resolve milliarcsecond-scale substructure in the lensed components. Correlated
substructure, such as a core and jet emission feature in each image, can
provide additional constraints to the model. Images resolved into two
subcomponents offer a total of ten constraints (eight image coordinates and
two flux density ratios), but the number of free parameters also increases by
two to account for a second set of unlensed source coordinates. For such
systems a sufficient number of constraints are available to study different
mass profiles. However, Monte Carlo results using sources consisting of two
closely-spaced subcomponents show that in order to determine the profile to
within 10\%, the galaxy position must be fixed by observations and the
subcomponents must be separated by several milliarcseconds (mas) in the lensed
images. 

Several radio lens systems are currently known to have image substructure. For
example, QSO 0957+561 (Barkana et al.\ 1999) features unusually rich jet
emission consisting of several correlated knots, which has allowed for the
mass profile of the lensing galaxy to be determined to 5--10\%. Each image of
CLASS B0218+357 (Biggs et al.\ 1999) is resolved into two subcomponents
separated by only $\sim 1$ mas, and the position of the lens galaxy is very
poorly known. CLASS B2319+051 (Marlow et al.\ 2000) has well-resolved
substructure, but thus far the lensed images have not been detected in the
optical, so the relative position of the lensing galaxy must be left as a free
parameter in the modeling process. The model is further complicated by the
presence of a nearly shear contributor at a different redshift than that of
the primary galaxy (Lubin et al.\ 2000).

\subsection{Four-image systems}

The situation with four-image lens systems is greatly improved, due to the
fact that the images probe the mass distribution at four distinct
locations. Image positions and flux density ratios provide a total of 11
constraints to the lens model, which allows the mass profile to be studied
directly. Monte Carlo simulations using quads produced with power-law
deflectors show that modeling can recover the true mass profile to within a
percent in $\gamma$, while strongly excluding neighboring profiles.
Therefore, quads lensed by an isolated galaxy offer the best possibility of
precisely measuring the Hubble constant through gravitational lensing.

Unfortunately, a very large fraction of known four-image lens systems have
lensing potentials that are significantly more complicated than that of a
single galaxy, which again leads to large uncertainties on both $H_o$ and the
profile. This is particularly true of the radio quads. Of the four-image
systems in the CLASS sample, two have multiple galaxies inside the Einstein
ring (B1359+154, Rusin et al.\ 2000; B1608+656, Koopmans and Fassnacht), and
two more appear to be significantly perturbed by strong external shear
contributors (B2045+265, Fassnacht et al.\ 1999; B1422+231, Kormann, Schneider
and Bartelmann 1994b). Two additional systems are lensed by a relatively
isolated galaxy (B0712+472, Jackson et al.\ 1998; B1933+507, Sykes et al.\
1998) but have thus far not shown themselves to be variable, which would
prevent the measurement of time delays. Finally, B1555+375 (Marlow et al.\
1999) has an inverted radio spectrum and is likely to be variable, but the
very small image separation ($\sim 420$ mas) makes a standard VLA monitoring
program infeasible. It is therefore questionable whether any of the currently
known set of four-image lens systems will provide tight constraints on
$H_o$. A subset of the systems may still be useful in studying the mass
profile, and advanced modeling of these lenses will be presented in a future
paper (Rusin 2000).

\section{Turning the problem around}

Gravitational lensing now competes with new methods seeking to measure the
Hubble constant. One technique in particular, combining the power spectrum of
the cosmic microwave background (CMB) with data from large galaxy redshift
surveys, has the potential to determine $H_o$ to within a few percent
(Eisenstein, Hu and Tegmark 1999) by the middle of the decade. Gravitational
lensing is unlikely to match this precision, due to aforementioned problems
with constraining the galaxy mass profiles. Witt, Mao and Keeton (2000) point
out that the uncertainties can be beaten down by measurements of $H_o$ in a
number of different lens systems. However, if the typical uncertainty is
$20\%$ on any particular measurement (as is currently the case for most
lenses), time delays for an unreasonably large number of systems ($\sim 
50$) would be required to reduce the overall error down to a competitive
level. This argument also assumes that the typical mass profile for lensing
galaxies corresponds to some known value (such as isothermal) which is input
to the lens models. If the wrong value is assumed, the derived Hubble
constants will converge to a systematically incorrect result.

Should the Hubble constant be sufficiently well determined by some other
means, however, the measured time delays become a direct and powerful
constraint on the slope of the mass profile. This astrophysical application of
the time delay method has been largely ignored, in favor of the cosmological
one.  However, this approach may provide the best tool for studying the inner
part of galaxy mass profiles and directly testing the isothermal
assumption. Two-image systems will be particularly important to this process,
as the majority of doubles appear to be dominated by a single lensing galaxy,
which will greatly ease the analysis.

One's ability to constrain the galaxy mass profile using this method depends
on the accuracy to which the image positions, flux density ratio and time
delay, as well as the galaxy position, can be measured. Radio imaging can
determine image positions to within a fraction of a milliarcsecond, so this
contributes very little to the uncertainty. History has also shown that flux
density ratios and time delays can measured from light curves to within a few
percent, given some persistence and luck (Kundic et al.\ 1997). 

Since two-image systems require that the deflector position be fixed to
constrain the model, any uncertainty in the center of the galaxy surface
brightness distribution will translate into a spread of predicted time delays,
and therefore weaken constraints on the mass profile.  Typically the
brightness center of a lens galaxy can be determined to no better than $10-25$
mas using the Hubble Space Telescope.\footnote{See the CASTLES webpage,
http://cfa-www.harvard.edu/castles/, for imaging data on most of the currently
known gravitational lenses.}  For a lens system with an image separation of
$\sim$ 1 -- 1.5 arcseconds, this translates to an uncertainty of $\sim 10\%$
in the predicted time delays when all the remaining galaxy parameters are
allowed to float. The error can be reduced if the position angle of the mass
model is constrained to be compatible with that of the observed surface
brightness distribution of the lens galaxy. This position angle may be a good
modeling constraint, as it is reasonable to believe that the light and mass
distributions are similarly oriented.

If both the Hubble constant and time delay are known to within a few percent,
the lens model is required to reproduce $\Delta t_{pred} = h \Delta t_{meas}$
(where $\Delta t_{pred}$ is the predicted delay for $h=1$) to within $\sim
5\%$. Since Fig.\ 2 shows that $\Delta t_{pred} \propto \gamma$ for the
parameter range of interest, the mass profile would be measured to this
accuracy, assuming the position of the galaxy is perfectly known. Including
the uncertainty in the galaxy position but using the orientation of the
surface brightness distribution as a constraint will likely increase the error
in the derived mass profile to $\Delta \gamma = 0.05 - 0.1$ for most two-image
lens systems.

\section{Discussion}

Gravitational lensing has already contributed to our knowledge of both galaxy
mass distributions and the Hubble constant. The profiles of several lens
galaxies have been constrained at the $10-20\%$ level, while the ensemble of
lenses with measured time delays may have determined the Hubble constant to
$\sim 10\%$ (Koopmans and Fassnacht 1999). In this age of precision
cosmology and astrophysics, however, new techniques and observations promise
very accurate measurements of a wide range of quantities. It is therefore
necessary to reassess the goals and merits of more familiar methods. 

The traditional application of gravitational lensing seeks to measure the
Hubble constant, and requires that the mass distribution of the lensing galaxy
be modeled independently of the absolute time delays. Uncertainties in the
mass profile dominate the error budget in $H_o$ for every system
in which time delays are currently measured. At the same time, lensing must
compete with new techniques that have the potential to determine the Hubble
constant to within a few percent. 

The time delay method, however, can be elegantly inverted to provide powerful
constraints on galaxy mass profiles, once the Hubble constant has been
accurately measured by some other means. Whereas highly constrained lens
systems, such as quads or doubles with well-separated substructure, would be
needed to derive $H_o$ independent of any assumptions regarding the mass
profile for a given galaxy, time delay determinations in almost any simple
two-image system can make interesting statements about the mass distribution
present. In time this approach may provide the best tool for probing the inner
few $h^{-1}$ kpc in elliptical galaxies, and will nicely complement weak
lensing techniques that are sensitive to the mass profile at large radius
(Fischer et al.\ 2000). Because photometric studies by the CASTLES
collaboration (Kochanek et al.\ 2000) have found that lensing galaxies appear
to be consistent with an ordinary population of passively evolving
ellipticals, the results derived from lensing can be applied to typical
galaxies at intermediate redshift. In conclusion, the precision studies of
galaxy mass profiles offered by lensing suggest that the most significant
results from Refsdal's method may be astrophysical in nature, rather than
cosmological.

\acknowledgements

I would like to thank Max Tegmark for his careful reading of this manuscript,
and Chung-Pei Ma and Dan Marlow for many interesting conversations on the
topic of mass profiles in gravitational lensing. This research was partially
funded by NASA grant NAG5-6034.

\clearpage

\begin{figure*}
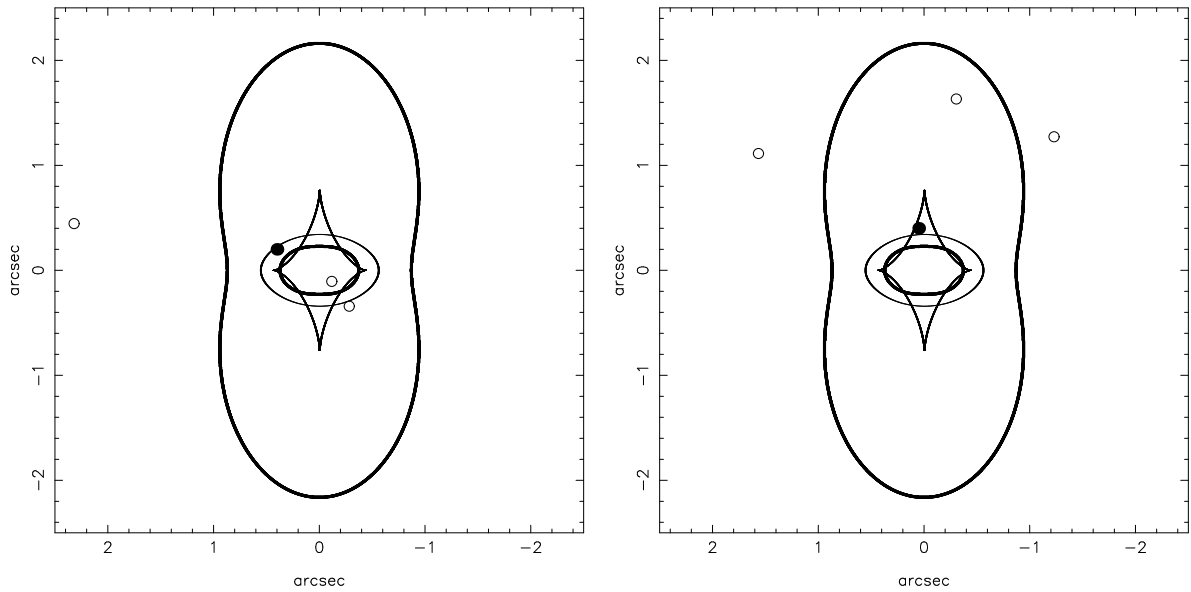

\begin{tabular}{c c}
\psfig{file=sple_fold.ps,width=3in} & 
\psfig{file=sple_cusp.ps,width=3in}\\
\end{tabular}
\figurenum{1}
\caption{Two classes of triple-imaging. (a) Left: Fold imaging. (b) Right:
Naked-cusp imaging. The critical curves are shown as dark lines and the
caustics are shown as light lines. The position of the unlensed source is
marked with a filled circle. The positions of the lensed images are marked
with open circles. The model shown is for $f = 0.6$, $\gamma = 0.6$.}
\end{figure*}

\clearpage

\begin{figure*}
\psfig{file=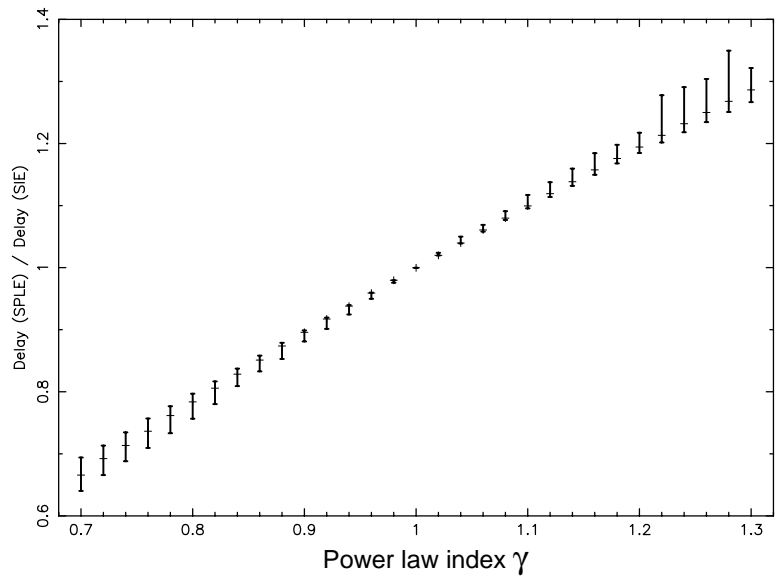,width=4in} 
\figurenum{2}
\caption{The variation of predicted time delays with mass profile for
two-image gravitational lens systems. Error bars indicate the range including
95\% of the Monte Carlo results.}
\end{figure*}

\end{document}